\documentclass[dvips]{acta}
\usepackage{longtable}
\usepackage{lscape}
\usepackage{epsfig}
\usepackage{amssymb}
\usepackage{amsmath}
\usepackage{url}
\usepackage{algorithmic}

\newcommand\actaa{\Acta}
\newcommand\apj{\ApJ}
\newcommand\apjs{\ApJS}
\newcommand\mnras{\MNRAS}

\SetPages{0}{0}

\SetVol{66}{2016}

\begin{document}

\begin{Titlepage}
  \Title{Analysis of photometric uncertainties in the OGLE-IV Galactic Bulge microlensing survey data\footnote{Based on observations obtained with the 1.3-m Warsaw telescope at the Las Campanas Observatory operated by the Carnegie Institution for Science.}}
\Author{
 J. S~k~o~w~r~o~n$^1$, 
 A. U~d~a~l~s~k~i$^1$, 
 S. K~o~z~{\l}~o~w~s~k~i$^1$,
 M.K. S~z~y~m~a~{\'n}~s~k~i$^1$, 
 P. M~r~{\'o}~z$^1$,
 {\L}. W~y~r~z~y~k~o~w~s~k~i$^1$,
 R. P~o~l~e~s~k~i$^{1,2}$,
 P. P~i~e~t~r~u~k~o~w~i~c~z$^1$,
 K. U~l~a~c~z~y~k$^1$,
 M. P~a~w~l~a~k$^1$,
and
I. S~o~s~z~y~{\'n}~s~k~i$^1$}
{$^1$Warsaw University Observatory, Al. Ujazdowskie 4, 00-478 Warszawa, Poland\\
e-mail:jskowron@astrouw.edu.pl\\
$^2$Department of Astronomy, Ohio State University, 140 West 18th Avenue, Columbus, OH 43210, USA}

\Received{March 13, 2016}
\end{Titlepage}

\Abstract{We present a statistical assessment of both, observed and reported, 
photometric uncertainties in the OGLE-IV Galactic bulge microlensing survey data. 
This dataset is widely used for the detection of variable stars, transient objects, 
discovery of microlensing events, and characterization of the exo-planetary systems.
Large collections of RR Lyrae stars and Cepheids discovered 
by the OGLE project toward the Galactic bulge provide light curves 
based on this dataset. 
We describe the method of analysis, and provide the procedure, 
which can be used to update preliminary photometric uncertainties,
provided with the light curves,
to the ones reflecting the actual observed scatter at a given magnitude
and for a given CCD detector of the OGLE-IV camera.
This is of key importance for data modeling, in particular, for the correct
estimation of the goodness of fit.}{surveys -- techniques: photometric -- methods: data analysis --  instrumentation: detectors -- stars: variables: general -- gravitational lensing: micro}

\section{Introduction}
A fundamental part of any scientific inference is an assessment of 
the input data uncertainties. 
Any following results need to have the realm of validity specified.
Also, any subsequent model parameters should have the uncertainties stated,
but these in turn strongly depend on the level of trust we had in the input data.

The photometric studies of transient objects rely on our ability
to firmly define the level of variability we trust is real.
It is important to differentiate any true signal from the observational noise 
(\eg Udalski \etal 1994, Wyrzykowski \etal 2014, Gould \etal 2015).
On the other hand, most of the studies of periodic variable stars 
do not depend on the well-characterized uncertainties. These periodic
signal can be found from the periodograms constructed only from the times and
the flux measurements 
(\eg Pojma{\'n}ski 2002, Soszy{\'n}ski \etal 2014).
However,
for very detailed studies of periodic stars (asteroseismology) 
the robust knowledge of levels of the photometric uncertainties is 
important 
(\eg Smolec \& Sniegowska 2016).
Additionally, accurate modeling of eclipsing systems can benefit from 
good estimation of the
photometric uncertainties 
(\eg Pilecki \etal 2013).
This is true 
especially for those cases where the brightness of object varies 
significantly ($\gtrsim1$~mag) since the relative weighting of the uncertainties
varies between the magnitude ranges.

The OGLE project is a large-scale sky survey for variability. Its fourth
phase 
(OGLE-IV, Udalski \etal 2015)
is under way and is regularly monitoring
650 square degrees of the Magellanic System (mainly for variables, transients, supernovae, etc.), 
130 square degrees of the Galactic bulge (microlensing, variables, novae) 
and 2200 square degrees of the Milky Way (variable stars, transient objects).
The OGLE-IV survey uses the 1.3-m Warsaw telescope at the Las Campanas Observatory
(of the Carnegie Institution for Science) equipped with the 32-CCD-chip mosaic
camera with the field of view of 1.4 square degrees. Consult 
Udalski \etal (2015)
for the detailed description of the project.

One of the key OGLE programs is monitoring of the Galactic bulge 
for the stellar variability with the cadence of 18 minutes to a couple of days. 
The observing season starts in February and lasts for nine months every year. 
Since 1994, the semi-automated system, {\it Early Warning System} (EWS), has been implemented to discover
the on-going microlensing events candidates 
(Udalski \etal 1994).

The light curves of the discovered microlensing events candidates are presented on the
web page \url{http://ogle.astrouw.edu.pl/ogle4/ews/ews.html} 
and are updated daily. 
These light curves are widely used by community in order to guide the follow-up efforts
and to facilitate the real-time modeling and anomaly detection 
(Gould \etal 2015, Dominik \etal 2008, Bozza \etal 2012, to mention a few).
Typically, the on-line data provide under-estimated instrumental uncertainties and it is a common
practice to rescale them with the use of the best-fit microlensing model, in order to force 
the $\chi^2$ per degree of freedom close to the unity 
(\eg Skowron \etal 2015).

The under-estimation of the uncertainties comes from two main sources. First,
the Difference Image Analysis (DIA) pipeline  
calculates the
expected uncertainty by propagation of the photon noise 
from the science frame onto the final measurement   
(Wo{\'z}niak \etal 2000, Wo{\'z}niak 2000).
Since it is hard to reach the theoretical photon-noise levels with the real-world data,
typical measurements have a slightly larger noise. Additionally, the reference image
for the subtraction
is regarded as noiseless. This is not strictly the case, especially 
if it is composed from the limited number of frames.
Second, the assumed point spread function (PSF) might not 
fully model the observed shapes of the stellar profiles 
(Wo{\'z}niak 2000).
This has a dominant effect at the bright end, where the uncertainties from the 
photon noise are far smaller than the uncertainties reported from the insufficient
model of the PSF 
(\eg Fig.~3 of Wo{\'z}niak 2000, and Fig.~2 of Wyrzykowski \etal 2009).

Due to the significant changes in magnitude, the analysis of microlensing events, 
detached eclipsing binaries, and cataclysmic variables is most susceptible to
any problems with the uncertainties of individual measurements. There is no single factor
which one can use in order to rescale the errorbars, and, that would be valid for 
the whole span of observed  magnitudes.
Hence, there is a need for a robust, empirical model of uncertainties and
the procedure that would correct their values at any magnitude.

It this paper we analyze 5.4 years of data gathered since June 29, 2010
until November 8, 2015 (JD=2455377--2457335) 
in the 85 frequently visited fields of the microlensing survey 
toward the Galactic bulge (see Section~2).
We present the analysis of the typical scatter observed in the photometric time series
of all stars and compare it to the reported errorbars in the OGLE-IV light curves.
We identify additional observational effects that have an impact on the observed scatter
and which are not fully taken into account within the routinely reported error bars (Sections~3 and~4).
We develop a series of functional forms to model the behavior of the true scatter 
across the whole magnitude range 
(Section~5).
These forms depend on the particular detector, typical observing conditions at the site and the 
details of the photometric pipeline.
We provide the procedure and all required coefficients to update photometric 
uncertainties of every measurement of the objects within this observational 
dataset (Section~6 and Appendix).

\Section{Data}
\label{sec:data}

\begin{figure}[htb]
  \includegraphics[width=\textwidth]{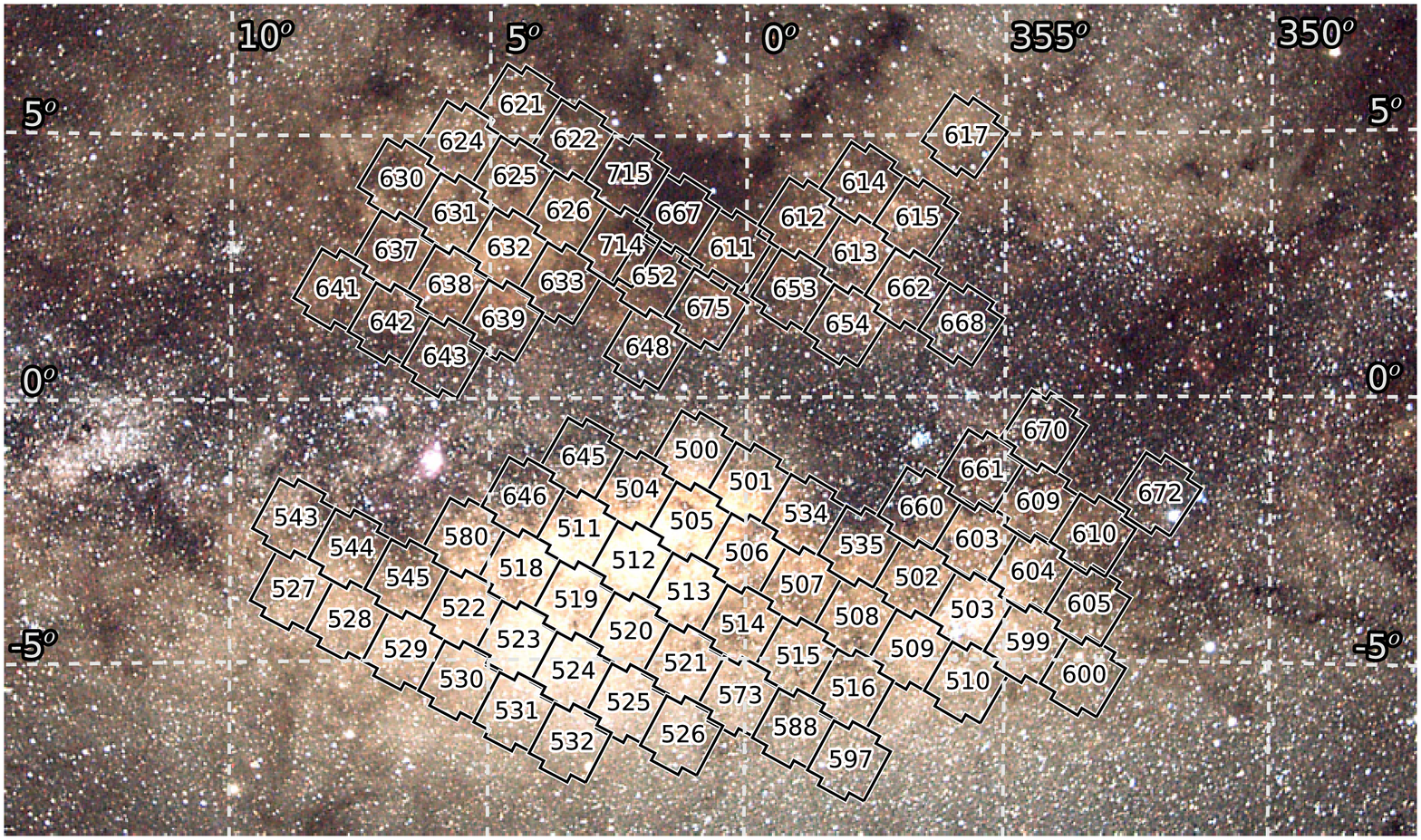}
  \FigCap{Fields of the microlensing survey within the OGLE-IV project.\label{fig:fields}}
\end{figure}

Fig.~\ref{fig:fields} shows all fields that are monitored with the 
current microlensing survey, which is a part of OGLE-IV program since 2010. 
For the details of the observing strategy and location of all OGLE-IV
fields toward the bulge see 
Udalski \etal (2015)
or consult the web page \url{http://ogle.astrouw.edu.pl/sky/ogle4-BLG/}.

The photometric database of the OGLE-IV project contains observation times (HJD), magnitudes and
uncertainties for all objects initially identified in the reference images.
The observations are performed in the $I$-band and the $V$-band. The majority of measurements
are performed with the former and between 300 and 13\,000 epochs have already 
been collected for every microlensing field. 
These data are used for transients detection and searches for the periodic stars.
The $V$-band observations have smaller
cadence and between 10 and 120 have been acquired in each field with the goal to help characterize
the discovered stars, microlensing events and to produce the Color Magnitude Diagrams (CMDs). 
Exposure times are 100~s and 150~s for the $I$- and $V$-band, respectively.

The OGLE-IV camera consists of 32 CCD detectors. Each has its own sensitivity,
gain and read-out noise. The gains are fine-tuned in such a way, that 
the 27.7~mag star would typically register 1 count
within a 100~s exposure.
Each observing field (\eg BLG672) are naturally 
divided into 32 subfields (chips) and referenced by adding a detector number (from 01 to 32)
to the name of the field 
(\eg BLG672.08).

\Section{Model of the Pipeline-Reported Error Bars}
\label{sec:errorbars}

All collected images are reduced with OGLE photometric pipeline based on the DIA photometry technique
(Wo{\'z}niak 2000).
Each photometric data point has an uncertainty evaluated by the propagation 
of the photon noise estimated for each pixel of the image through the 
linear least-squares with the flat-fielding uncertainty added in the quadrature.
The reference image is treated as noiseless.

\begin{figure}[htb]
  \includegraphics[width=\textwidth]{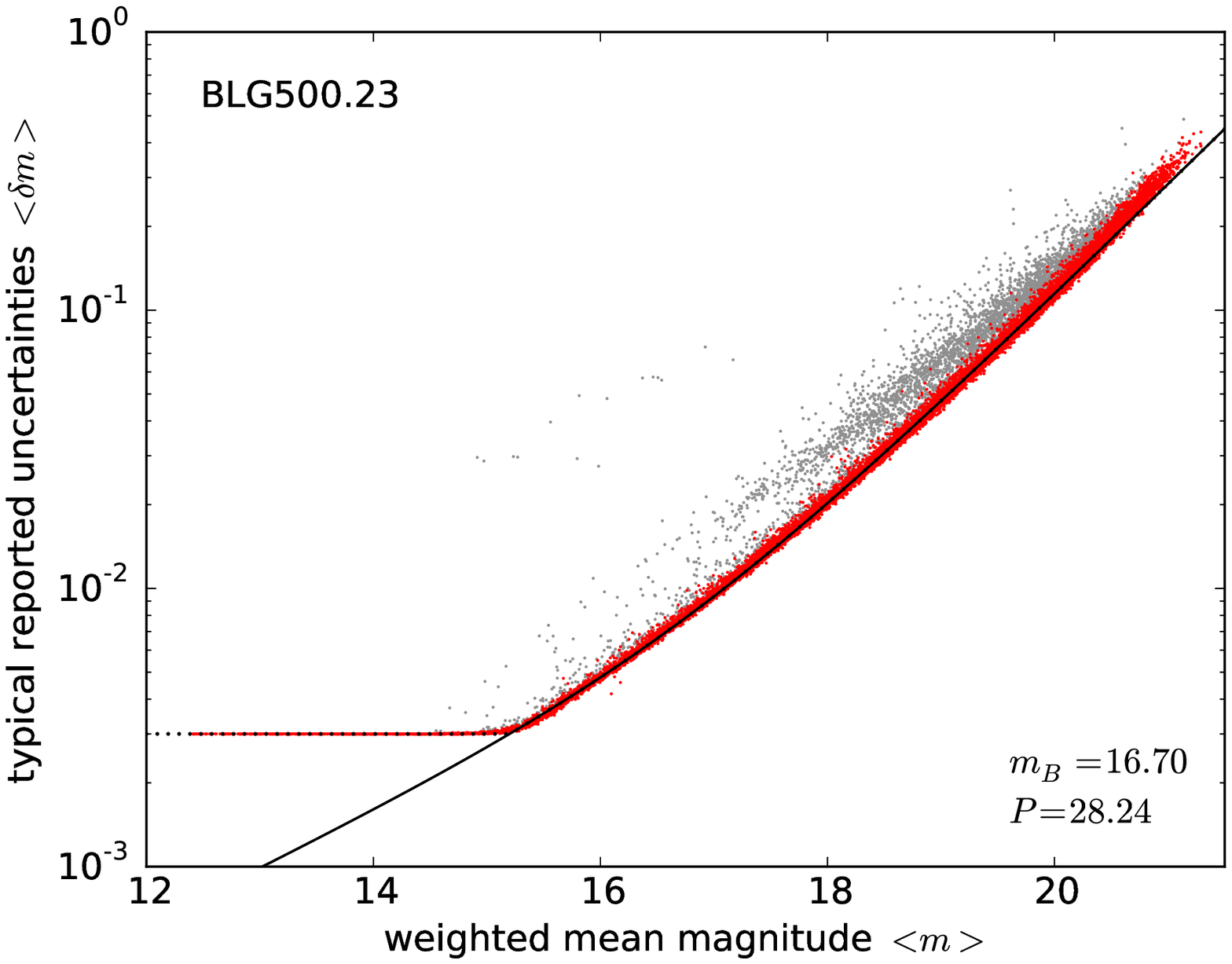}
  \FigCap{Mean reported uncertainties \vs the mean magnitude of a star (red points) in the $I$-band. Each point represents the light curve of an individual star in the field. The data for the BLG500.23 field are shown for illustration. The exact behavior changes slightly depending on the field and on the CCD detector. The solid line shows the photon-noise model estimated by the fit to the typical uncertainties. The uncertainties with the values below 3 mmag, as not reliable, are clipped in the database to 3 mmag. The gray points represent the rejected, spurious objects -- spikes near the bright stars and reflections at the edges of the camera.\label{fig:err}}
\end{figure}

Fig.~\ref{fig:err} presents the the typical uncertainties reported
for the $I$-band light curves. Each dot represents one object for which
the mean weighted magnitude and the mean uncertainty of a random sample of 100 measurements
was calculated. We show objects located in the field BLG500 and measured with the CCD
detector no. 23 as an illustration. The exact values of the uncertainties, for a given brightness, vary 
by a few percent depending on the detector ($\approx7\%$) and the field ($\approx3\%$). 
All uncertainties equal to or below 3~mmag are treated as unrealistic. 
They are clipped to exactly 3~mmag in the photometric database
and in the on-line light curves. 

For each field and detector pair we characterize the typical uncertainties with a simple 
 two-parameter model consisting of a Poisson noise contribution from the object and from the background. That is
fitted to the empirical data -- as presented in Fig.~\ref{fig:err}.
The signal from the object of the mean brightness, $m$, is calculated with the Eq.~\ref{eq:flux},
and the signal from the background is a parameter of the model ($m_B$).
The second parameter is a number that translates the brightness into the photon count ($P$) in the following way:
\begin{gather}
  \label{eq:flux}
  F = 10^{0.4 (P - m)} \\
  B = 10^{0.4 (P - m_B)}.
\end{gather}
Then, we estimate the noise as a square-root of the signal
\begin{equation}
  \Delta F = \sqrt{F + B}
\end{equation}
and translate it into the magnitude scale with
\begin{equation}
  \Delta m = 2.5/\ln{10} \times \Delta F/F.
\end{equation}

There are many more faint stars than the bright ones. In order for the fit to be more stable, we bin the sample in
50-100 bins in magnitude (depending on the number of stars in the given subfield) and calculate the median of
the typical uncertainties within each bin. The input data consist of about a 100 pairs of magnitude and uncertainty  $(m_i, \delta m_i)$ to 
which the noise model is fitted with the least-squares method.
We perform the fit in the magnitude range from $\approx15.3$~mag, where the errorbar values are meaningful, to 20.5~mag, above which the mean
brightness measurement might be less reliable.
The noise model fitted to the data ($\Delta m(m_i) = \delta m_i$) in the field BLG500.23 is shown with a solid line in Fig.~\ref{fig:err}.
In the Appendix, we list the fitted values of $P$ and $m_B$ for each pair of the field and the detector in the data set.

\Section{Measurement of the Observational Scatter}
\label{sec:scatter}

\begin{figure}[htb]
  \includegraphics[width=\textwidth]{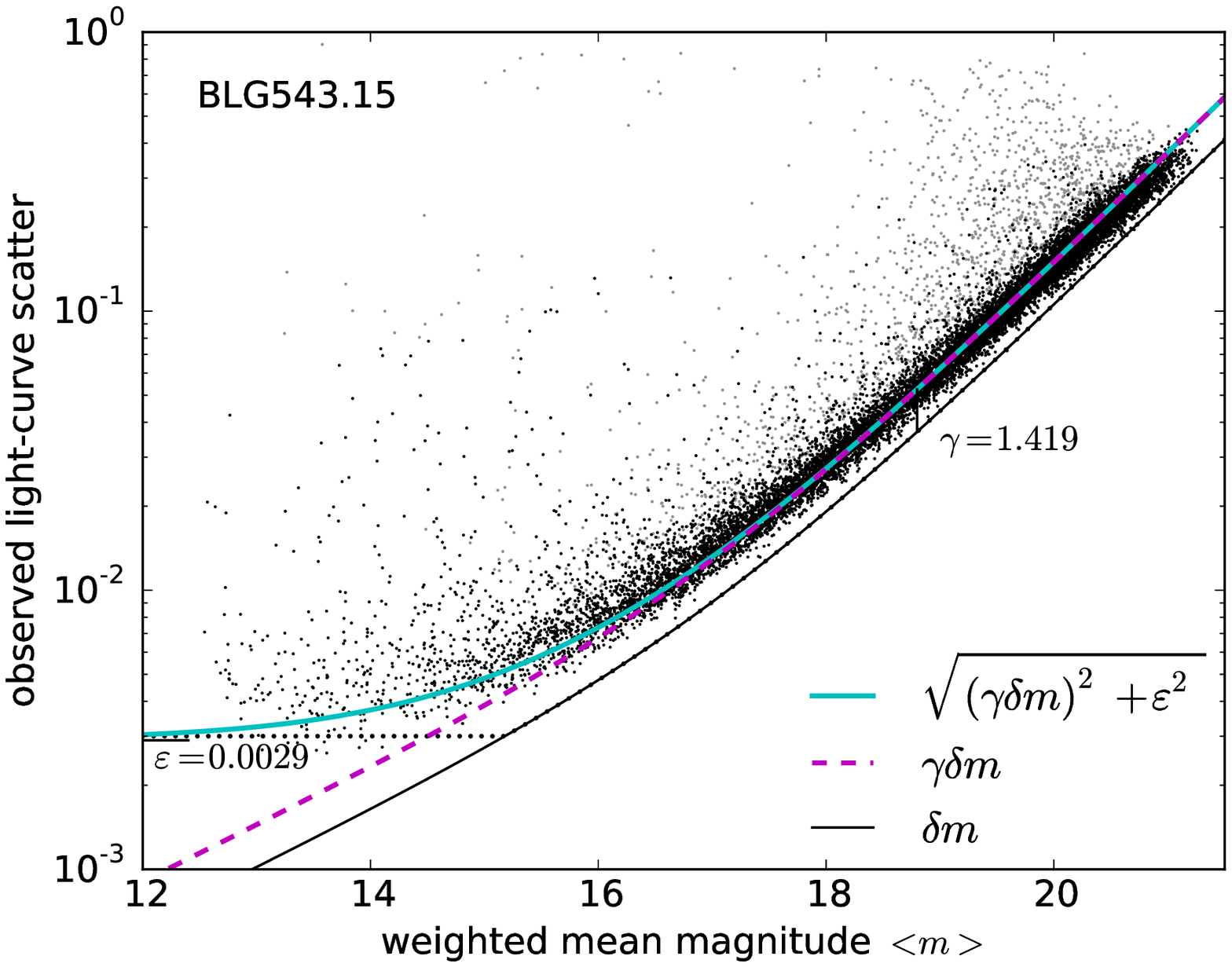}
  \FigCap{Mean root mean square light-curve scatter as a function of the mean magnitude 
  of a star (black points) in the $I$-band. 
  Each point represents the light curve of an individual star in the field. 
  The data for the BLG543.15 field are shown for illustration. 
  The correction parameters slightly change, depending on the field and on the CCD detector. 
  The black solid and dotted lines show the photon-noise model and typical reported 
  uncertainties as in the Fig.~\ref{fig:err}. 
  The magenta dashed line represents the noise model multiplied by a factor $\gamma$ in
  order to match it to the observed scatter. 
  The cyan solid line shows the same model, but with the additional constant error ($\epsilon$)
  added in quadrature. This is the empirical term that dominates for the bright stars 
  ($\sqrt{(\gamma \delta m)^2 + \epsilon^2}$).
  The gray points represent the rejected, spurious objects -- spikes near 
  the bright stars and reflections at the edges of the camera.\label{fig:rms}}
\end{figure}

In order to characterize the real observational uncertainty as a function 
of an object magnitude in a given field and given detector we analyze the light curves of all stars in this field.
For each object we choose 100 random measurements from its time-series photometry 
and calculate the weighted mean magnitude and a root-mean-square scatter. 
Such data are presented in the Fig.~\ref{fig:rms} for a BLG543.15 field, where each dot represents a single object.
The model of the typical uncertainties reported by the pipeline is presented as a black solid line.
It is clear from the figure, that the uncertainties are under-estimated as the observed scatter for the majority of objects
is larger than the noise model.

In the same fashion as in the previous Section, we divide the sample into 50-100 magnitude bins and find the value of scatter
that is representative for every bin.
We expect some of the stars to be variable, hence, we ignore 20\% of points showing the largest scatter in the given bin and calculate the median of the remaining points.
Then, we can fit for the scaling parameter ($\gamma$) which shifts 
the noise model to the value of the observed scatter (see Fig.~\ref{fig:rms}).

For the stars brighter than $\sim$16th magnitude we see that there is an additional source of scatter that is
not due to the photon noise. The scatter is consistent with the constant error level of a few 
millimagnitudes.
This effect have already been found by various researchers -- see Fig.~3 of 
Wo{\'z}niak (2000)
or Fig.~3 of 
Sumi \etal (2003).
The constant error floor for the bright measurements is also 
routinely used in the process of rescaling error bars in the microlensing events
analyzes 
(\eg Wyrzykowski \etal 2009, Skowron \etal 2016).
Alard \& Lupton (1998) and Wo{\'z}niak (2000)
suggest that the potential source of 
additional scatter is an inability of the mathematical PSF model to 
perfectly approximate the real-life seeing. The small deviations are not
visible for the faint and moderate-brightness stars; this is noticeable only where the signal 
is strong and the photon noise small.

For each subfield we introduce the additional parameter, $\epsilon$, aimed to model 
the level of the fixed error floor in the observed light curves. Then, for
each measurement with the formal uncertainty, $\delta m$, that was performed 
with a given detector and in a given field, we can estimate the more optimal value
of the uncertainty with $\sqrt{(\gamma \delta m)^2 + \epsilon^2}$.
The values of $\gamma$ and $\epsilon$ for every subfield are tabulated in the Appendix.
Fig.~\ref{fig:rms} presents this two-parameter model with respect to the observed light-curve scatter.

\Section{Model of the Observed Scatter}
\label{sec:model}

For the measurements fainter than $\approx15.2$~mag the values of errorbars 
provided in the light curves are typically greater than 3 mmag.
These values carry 
some information about the conditions the image was taken under.
The image parameters, like seeing and background have straight
impact on the number of pixels that enters into the photon-noise estimation
as well as the noise estimated within those pixels.
For these measurements we employ the formula 
\begin{equation}
  \delta m_{i,\rm new} = \sqrt{(\gamma \delta m_i)^2 + \epsilon^2}\qquad (\text{if }\delta m_i>0.003),
  \label{eq:gammeps}
\end{equation}
where $m_i$ is the original uncertainty. 
Note, that the $P$ and $m_B$
parameters introduced in Section~3 model only the median
noise under typical seeing and background conditions at the survey site. 
In contrast, Eq.~\ref{eq:gammeps} is designed to conserve the information
about the relative quality of each measurement.

\begin{figure}[htb]
  \includegraphics[width=\textwidth]{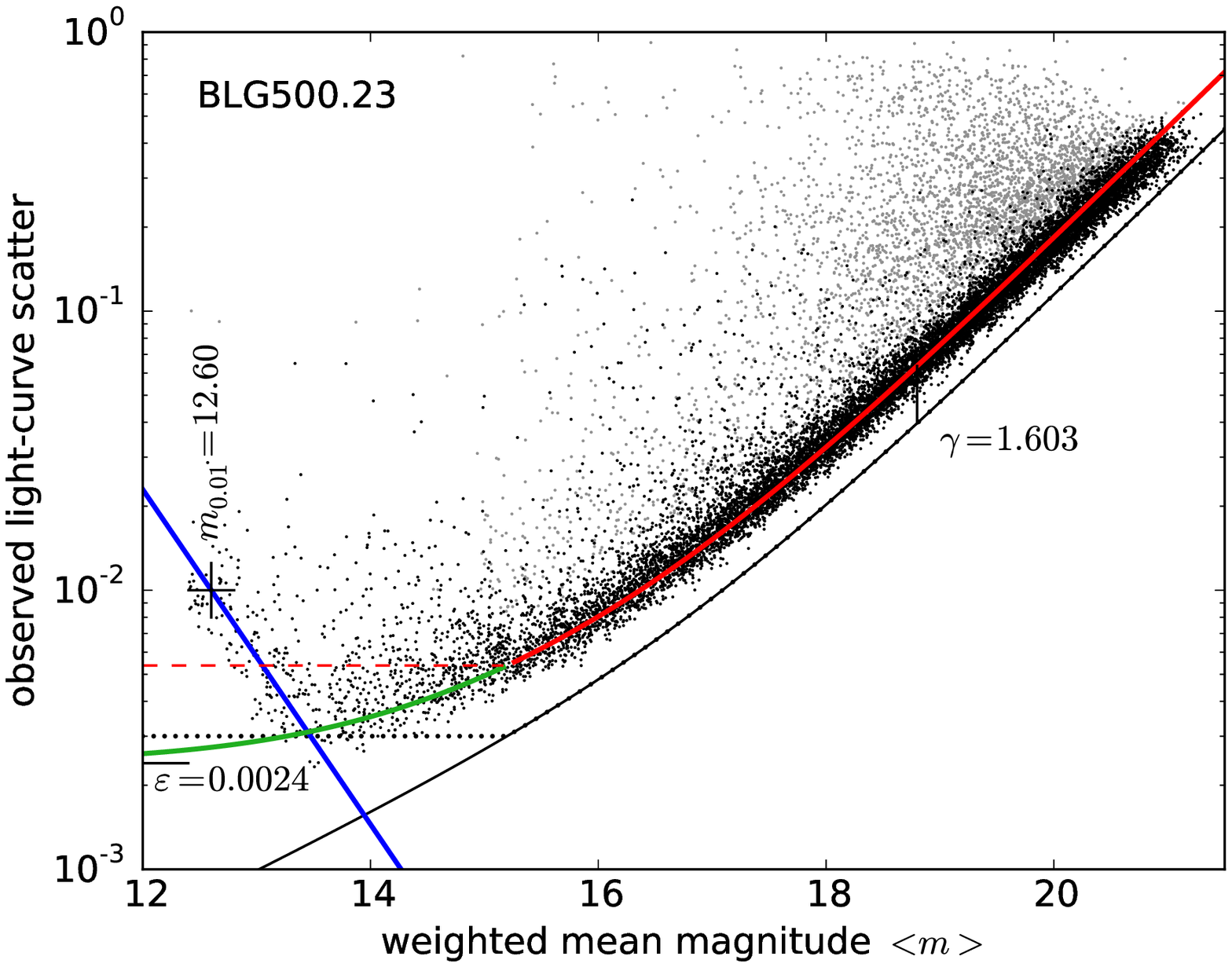}
  \FigCap{Similar to Fig.~\ref{fig:rms} but with the final models of the scatter plotted.
  The blue, green and red solid lines represent three models of the light-curve scatter 
  in the three, separate magnitude regimes: very bright (where non-linearity of the 
  detector dominates), bright (where uncertainties are estimated from the photon-noise model), 
  and remaining (where the majority of stars are and the canonical scaling of the 
  uncertainties by $\sqrt{(\gamma \delta m)^2+\epsilon^2}$ is possible).\label{fig:rmsall}}
\end{figure}

All the uncertainties with values below 0.003 mag are not reliable, 
thus, we estimate the most likely 
uncertainty of such a measurement by taking into account only its
brightness and comparing it to the typical scatter shown by the constant stars
at the similar brightness.
We use the model of the noise, 
as described in Sec.~3. We rescale it by $\gamma$ and add the
error floor $\epsilon$ in quadrature in order to better approximate the observational
scatter for the bright stars.
This averaged model for bright stars is presented with the green solid line in Fig.~\ref{fig:rmsall}.

We note, that there is, however, a yet another source of scatter that is evident for the
very brightest stars ($I \lesssim13$~mag). There is a very fast rising trend of scatter
with the increasing brightness. This was not present in the OGLE-II data
(see Fig. 2 of Wyrzykowski \etal 2009)
nor in the OGLE-III data 
(\cf Wyrzykowski \etal 2011).
In the laboratory studies, the CCD detectors of the OGLE-IV camera do show slight
level of non-linear response ($\approx0.1{-}0.3\%$) for pixels with 100k electrons.
This corresponds to about 52k--62k counts depending on the gain of a particular detector.
The limit above which the pixel is not measured by the pipeline is 65\,535 counts. 
This might be an explanation for a very fast raise of uncertainties for extremely bright objects
where the central pixels are close to this limit.

The fast rise of uncertainties is consistent with the scatter of flux being proportional to $\sim F^{5/2}$
(where $F$ is the observed flux).
This is an empirical approximation, however, it is in an good agreement with the data from the various 
fields and detectors.
We parametrize this effect by providing the magnitude value, $m_{0.01}$, at which the expected scatter for constant stars
is equal to 0.01 mag. Since the slope is fixed, we can provide a one-parameter 
formula for the uncertainty due to this effect for every photometric measurement $m_i$ in the form:
\begin{equation}
  \Delta m_{nonlinear,i} = 0.01 \times 10^{0.6 (m_{0.01} - m_i)}
  \label{eq:nonl}
\end{equation}

During the estimation of the single-parameter $m_{0.01}$ we notice that it has very similar value for
the same CCD detector across multiple observing fields. This further strengthens the argument, 
that it is a feature of the particular detector.
Since the number of very bright stars is limited, we can merge data from various 
fields in order to robustly estimate the value of $m_{0.01}$ for any of 32 detectors.
In order not to overestimate the typical scatter for the very bright stars, instead of a median,
we choose $m_{0.01}$ in such a way that only 15 percent of points lie below the model described by Eq.~\ref{eq:nonl}.
This is done, because the distribution of scatter values form the brightest stars is not Gaussian, and 15th percentile
lies closer to the mode of the distribution than the mean or the median. 

With the better understanding of the non-linear behavior of CCD detectors and with additional information 
about conditions of a particular exposure (seeing, background) it is possible to better estimate magnitude and
uncertainty of a measurement. Here, we only provide a statistical formula that describes the typical
expected scatter from this effect and can be used with an absence of any additional information.

\Section{Correction Procedure for the Uncertainties}
\label{sec:corr}

Any light curve from the OGLE-IV real-time data analysis systems 
(Udalski 2008, Udalski \etal 2015, Mr{\'o}z \etal 2015b),
the on-going
microlensing event from the Early Warning System\footnote{\url{http://ogle.astrouw.edu.pl/ogle4/ews/ews.html}}
or the light curve from the early catalogs of variable stars based on the OGLE-IV Galactic bulge
data 
(\eg Poleski \etal 2011, Mr{\'o}z \etal 2015a),
can be corrected using the procedure described in this Section.
The uncertainties rescaled based on the statistical behavior of the neighboring constant stars can
facilitate the detailed modeling of the individual objects.

The full procedure is described in the next paragraph. The input to the procedure consist of two values:
the magnitude and the uncertainty of the individual measurement $(m_i, \delta m_i)$ from
the OGLE-IV light curve. One has to also know the observed field
and the particular detector the measurement was taken with. The Appendix lists
the five parameters for every field and CCD chip pair. These parameters are as follows:
the scale of the level of errorbar underestimation ($\gamma$),
the constant error floor, which dominates for the bright measurements ($\epsilon$),
the approximated counts-to-magnitudes conversion unit ($P$),
the value parameterizing the background noise ($m_B$),
and the magnitude at which the scatter from the non-linearity effect of 
the detector reaches 0.01 magnitudes ($m_{0.01}$).
The detailed meaning of each parameters are described in previous Sections.

The procedure is as follows:

\renewcommand{\algorithmicrequire}{\textbf{Input:}}
\renewcommand{\algorithmicensure}{\textbf{Output:}}
\renewcommand{\algorithmiccomment}[1]{-- #1}
\begin{algorithmic}
\REQUIRE $(m_i, \delta m_i)$ -- a photometric data point of interest
and $(\gamma, \epsilon, P, m_B, m_{0.01})$ -- the coefficients for the given FIELD.CHIP pair
  \IF{$\delta m_i$ > 0.003 mag}
    \STATE $\delta m_{{\rm new},i} = \sqrt{ (\gamma \delta m_i)^2 + \epsilon^2 }$
  \ELSE
  \STATE $\Delta m_{nonlinear} = 0.01 \times 10^{0.6 (m_{0.01} - m_i)}$ \COMMENT{model of non-linearity ($\sim F^{5/2}$)}
    \STATE $F = 10^{0.4 (P - m_i)}$ \COMMENT{approximate number of photons from the star}
    \STATE $B = 10^{0.4 (P - m_B)}$ \COMMENT{approximate number of background photons contributing to the noise}
    \STATE $\Delta F = \sqrt{F + B}$ \COMMENT{photon noise model}
    \STATE $\Delta m_{bright} = 2.5/\ln{10} \times \Delta F/F$ \COMMENT{estimated noise in magnitudes}
    \STATE $\delta m_{{\rm new},i} = max \left[ \delta m_i, \sqrt{ (\gamma \Delta m_{bright})^2 + \epsilon^2 }, \Delta m_{nonlinear} \right]$
  \ENDIF
\ENSURE $(m_i, \delta m_{{\rm new},i})$
\end{algorithmic}

The output from the procedure is the new value of the uncertainty, which goal is to provide the better estimation of the expected
accuracy of the measurement.

\Section{Conclusions}
We have performed statistical analysis of the typical light-curve scatter 
in the OGLE-IV Galactic bulge microlensing survey data.
The scatter for non-variable stars was compared to the 
mean uncertainty reported by the photometric pipeline at various magnitudes.
The amount of errorbar underestimation was measured for every field
and every detector and it is typically between 10\% and 70\%. 
The scaling values ($\gamma$) are presented in the Appendix. 

We found additional effects that should be taken into account
within the light-curve errorbars. Most importantly, the existence 
of a constant error floor, which affects the uncertainties of 
brighter measurements ($\lesssim15$~mag) and is parametrized by $\epsilon$
(typically $\approx2.8$~mmag).

The scatter of 12--13~mag stars is
much greater than the scatter of 13--14~mag stars.
This is a likely effect of the non-linearity of the CCD detectors for the very 
large signals.
We find that the rise of the scatter with the increasing flux of the star
can be described by the power law ($\sim F^{5/2}$) and we
measure this effect for every CCD detector of the OGLE-IV camera.

The detailed procedure for correcting the values of reported uncertainties
is provided in Section~6. This procedure can be applied to
any light curve in the discussed fields in order to facilitate the analysis
of variable stars and transient objects.

Finally, in the Appendix, we provide the values needed as an inputs to the above mentioned procedure.
These values are quoted for every subfield of the survey and were estimated 
from the analysis of light curves of all objects within the given region.

\Acknow{
  We would like to thank Profs.\ M. Kubiak and G. Pietrzy{\'n}ski, former
  members of the OGLE team, for their contribution to the collection of
  the OGLE photometric data over the past years.
  The OGLE project has received funding from the National Science Center, Poland,
grant MAESTRO 2014/14/A/ST9/00121 to Andrzej Udalski.}

\appendix
\Section{Error-bar correction coefficients for every FIELD.CHIP pair}
\label{sec:app}
Below we present the coefficients for the uncertainty-correction procedure 
described in Sec.~6.
The full on-line tables are available at WWW address: 

\url{http://ogle.astrouw.edu.pl/ogle4/errorbars/blg/}

and FTP address: 

\url{ftp://ftp.astrouw.edu.pl/ogle/ogle4/errorbars/}.

\begin{footnotesize}
\begin{longtable}{cccccc|cccccc}
\hline
CCD & $\gamma$ & $\epsilon$ & $P$ & $m_B$ & $m_{0.01}$ & CCD & $\gamma$ & $\epsilon$ & $P$ & $m_B$ & $m_{0.01}$\\
\hline
\multicolumn{12}{l}{BLG501 ($I$-band)} \\ 
.01 & 1.731 & 0.0030 & 28.215 & 16.217 & 12.491   &  .02 & 1.651 & 0.0030 & 28.233 & 16.317 & 12.354 \\ 
.03 & 1.663 & 0.0032 & 28.252 & 16.344 & 12.734   &  .04 & 1.696 & 0.0031 & 28.258 & 16.367 & 12.453 \\ 
.05 & 1.724 & 0.0028 & 28.268 & 16.437 & 12.495   &  .06 & 1.664 & 0.0028 & 28.262 & 16.547 & 12.524 \\ 
.07 & 1.686 & 0.0028 & 28.278 & 16.671 & 12.641   &  .08 & 1.756 & 0.0031 & 28.231 & 16.055 & 12.659 \\ 
.09 & 1.650 & 0.0028 & 28.241 & 16.234 & 12.557   &  .10 & 1.660 & 0.0030 & 28.254 & 16.246 & 12.581 \\ 
.11 & 1.677 & 0.0031 & 28.265 & 16.325 & 12.374   &  .12 & 1.769 & 0.0034 & 28.277 & 16.379 & 12.655 \\ 
.13 & 1.671 & 0.0031 & 28.278 & 16.457 & 12.323   &  .14 & 1.687 & 0.0025 & 28.296 & 16.526 & 12.367 \\ 
.15 & 1.627 & 0.0024 & 28.299 & 16.713 & 12.258   &  .16 & 1.754 & 0.0030 & 28.266 & 16.617 & 12.530 \\ 
.17 & 1.804 & 0.0029 & 28.228 & 16.080 & 12.488   &  .18 & 1.689 & 0.0031 & 28.230 & 16.218 & 12.800 \\ 
.19 & 1.738 & 0.0030 & 28.246 & 16.222 & 12.573   &  .20 & 1.722 & 0.0029 & 28.258 & 16.247 & 12.567 \\ 
.21 & 1.784 & 0.0037 & 28.280 & 16.316 & 12.487   &  .22 & 1.677 & 0.0029 & 28.289 & 16.442 & 12.421 \\ 
.23 & 1.669 & 0.0028 & 28.288 & 16.535 & 12.602   &  .24 & 1.706 & 0.0024 & 28.315 & 16.634 & 12.300 \\ 
.25 & 1.653 & 0.0027 & 28.284 & 16.698 & 12.400   &  .26 & 1.754 & 0.0028 & 28.247 & 16.275 & 12.370 \\ 
.27 & 1.735 & 0.0027 & 28.256 & 16.391 & 12.334   &  .28 & 1.706 & 0.0030 & 28.268 & 16.462 & 12.772 \\ 
.29 & 1.713 & 0.0031 & 28.273 & 16.459 & 12.660   &  .30 & 1.609 & 0.0029 & 28.287 & 16.687 & 12.361 \\ 
.31 & 1.615 & 0.0025 & 28.289 & 16.691 & 12.370   &  .32 & 1.474 & 0.0027 & 28.300 & 17.038 & 12.474 \\ 
\hline
\multicolumn{12}{l}{BLG505 ($I$-band)} \\ 
.01 & 1.575 & 0.0033 & 28.096 & 16.281 & 12.491   &  .02 & 1.480 & 0.0029 & 28.099 & 16.353 & 12.354 \\ 
.03 & 1.527 & 0.0033 & 28.111 & 16.315 & 12.734   &  .04 & 1.519 & 0.0030 & 28.106 & 16.342 & 12.453 \\ 
.05 & 1.541 & 0.0030 & 28.109 & 16.400 & 12.495   &  .06 & 1.526 & 0.0029 & 28.116 & 16.374 & 12.524 \\ 
.07 & 1.532 & 0.0032 & 28.104 & 16.348 & 12.641   &  .08 & 1.564 & 0.0039 & 28.095 & 16.266 & 12.659 \\ 
.09 & 1.456 & 0.0032 & 28.101 & 16.425 & 12.557   &  .10 & 1.489 & 0.0029 & 28.108 & 16.347 & 12.581 \\ 
.11 & 1.486 & 0.0030 & 28.124 & 16.372 & 12.374   &  .12 & 1.547 & 0.0037 & 28.122 & 16.398 & 12.655 \\ 
.13 & 1.495 & 0.0030 & 28.122 & 16.381 & 12.323   &  .14 & 1.500 & 0.0029 & 28.127 & 16.386 & 12.367 \\ 
.15 & 1.466 & 0.0026 & 28.127 & 16.395 & 12.258   &  .16 & 1.628 & 0.0032 & 28.113 & 16.186 & 12.530 \\ 
.17 & 1.653 & 0.0030 & 28.098 & 16.194 & 12.488   &  .18 & 1.563 & 0.0034 & 28.105 & 16.273 & 12.800 \\ 
.19 & 1.579 & 0.0026 & 28.104 & 16.223 & 12.573   &  .20 & 1.572 & 0.0029 & 28.117 & 16.266 & 12.567 \\ 
.21 & 1.581 & 0.0033 & 28.118 & 16.297 & 12.487   &  .22 & 1.486 & 0.0031 & 28.116 & 16.460 & 12.421 \\ 
.23 & 1.486 & 0.0030 & 28.114 & 16.452 & 12.602   &  .24 & 1.514 & 0.0029 & 28.139 & 16.371 & 12.300 \\ 
.25 & 1.550 & 0.0032 & 28.109 & 16.289 & 12.400   &  .26 & 1.611 & 0.0027 & 28.118 & 16.306 & 12.370 \\ 
.27 & 1.569 & 0.0026 & 28.110 & 16.359 & 12.334   &  .28 & 1.559 & 0.0032 & 28.112 & 16.373 & 12.772 \\ 
.29 & 1.549 & 0.0030 & 28.106 & 16.334 & 12.660   &  .30 & 1.423 & 0.0031 & 28.117 & 16.512 & 12.361 \\ 
.31 & 1.448 & 0.0028 & 28.117 & 16.433 & 12.370   &  .32 & 1.345 & 0.0037 & 28.140 & 16.694 & 12.474 \\ 
\hline
\multicolumn{12}{l}{BLG512 ($I$-band)} \\ 
.01 & 1.731 & 0.0026 & 28.256 & 16.273 & 12.491   &  .02 & 1.639 & 0.0024 & 28.264 & 16.356 & 12.354 \\ 
.03 & 1.661 & 0.0025 & 28.274 & 16.322 & 12.734   &  .04 & 1.702 & 0.0025 & 28.267 & 16.292 & 12.453 \\ 
.05 & 1.744 & 0.0023 & 28.275 & 16.321 & 12.495   &  .06 & 1.688 & 0.0023 & 28.268 & 16.371 & 12.524 \\ 
.07 & 1.708 & 0.0027 & 28.266 & 16.343 & 12.641   &  .08 & 1.703 & 0.0028 & 28.242 & 16.238 & 12.659 \\ 
.09 & 1.620 & 0.0025 & 28.260 & 16.365 & 12.557   &  .10 & 1.636 & 0.0025 & 28.267 & 16.329 & 12.581 \\ 
.11 & 1.651 & 0.0024 & 28.267 & 16.318 & 12.374   &  .12 & 1.760 & 0.0028 & 28.269 & 16.259 & 12.655 \\ 
.13 & 1.674 & 0.0024 & 28.269 & 16.299 & 12.323   &  .14 & 1.683 & 0.0024 & 28.274 & 16.280 & 12.367 \\ 
.15 & 1.649 & 0.0022 & 28.273 & 16.283 & 12.258   &  .16 & 1.835 & 0.0026 & 28.268 & 16.042 & 12.530 \\ 
.17 & 1.783 & 0.0027 & 28.241 & 16.283 & 12.488   &  .18 & 1.660 & 0.0025 & 28.259 & 16.369 & 12.800 \\ 
.19 & 1.730 & 0.0022 & 28.254 & 16.308 & 12.573   &  .20 & 1.707 & 0.0024 & 28.259 & 16.262 & 12.567 \\ 
.21 & 1.747 & 0.0027 & 28.257 & 16.230 & 12.487   &  .22 & 1.668 & 0.0024 & 28.261 & 16.263 & 12.421 \\ 
.23 & 1.660 & 0.0023 & 28.271 & 16.278 & 12.602   &  .24 & 1.672 & 0.0024 & 28.276 & 16.282 & 12.300 \\ 
.25 & 1.708 & 0.0027 & 28.265 & 16.180 & 12.400   &  .26 & 1.750 & 0.0021 & 28.251 & 16.347 & 12.370 \\ 
.27 & 1.726 & 0.0020 & 28.262 & 16.410 & 12.334   &  .28 & 1.683 & 0.0025 & 28.256 & 16.388 & 12.772 \\ 
.29 & 1.716 & 0.0023 & 28.254 & 16.277 & 12.660   &  .30 & 1.599 & 0.0024 & 28.254 & 16.389 & 12.361 \\ 
.31 & 1.600 & 0.0025 & 28.267 & 16.397 & 12.370   &  .32 & 1.486 & 0.0031 & 28.275 & 16.627 & 12.474 \\ 
\hline
\multicolumn{12}{l}{\ldots table abbreviated, see on-line material.} \\ 
\hline
\caption{%
The errorbar-correction coefficients for the $I$-band
data from the 1.3-m Warsaw Telescope made with the 
32-CCD-chip mosaic OGLE-IV camera.
The observing fields in the table are from the 
on-going microlensing survey toward
the Galactic bulge.
The full contents of this table is available 
on-line at \emph{ftp://ftp.astrouw.edu.pl/ogle/ogle4/errorbars/errcorr-OIV-BLG-I.dat}.
}
\end{longtable}
\end{footnotesize}
\begin{footnotesize}
\begin{longtable}{cccccc|cccccc}
\hline
CCD & $\gamma$ & $\epsilon$ & $P$ & $m_B$ & $m_{0.01}$ & CCD & $\gamma$ & $\epsilon$ & $P$ & $m_B$ & $m_{0.01}$\\
\hline
\multicolumn{12}{l}{BLG500 ($V$-band)} \\ 
.01 & 1.468 & 0.0032 & 28.838 & 18.097 & 13.286   &  .02 & 1.392 & 0.0026 & 28.852 & 18.398 & 13.108 \\ 
.03 & 1.435 & 0.0027 & 28.880 & 18.459 & 13.515   &  .04 & 1.453 & 0.0026 & 28.876 & 18.540 & 13.186 \\ 
.05 & 1.538 & 0.0027 & 28.872 & 18.652 & 13.186   &  .06 & 1.487 & 0.0023 & 28.881 & 18.685 & 13.241 \\ 
.07 & 1.580 & 0.0028 & 28.899 & 18.634 & 13.363   &  .08 & 1.410 & 0.0028 & 28.847 & 18.207 & 13.481 \\ 
.09 & 1.393 & 0.0024 & 28.864 & 18.354 & 13.330   &  .10 & 1.407 & 0.0025 & 28.870 & 18.458 & 13.310 \\ 
.11 & 1.441 & 0.0027 & 28.880 & 18.558 & 13.068   &  .12 & 1.483 & 0.0032 & 28.872 & 18.536 & 13.383 \\ 
.13 & 1.439 & 0.0027 & 28.881 & 18.598 & 13.003   &  .14 & 1.469 & 0.0020 & 28.890 & 18.612 & 13.041 \\ 
.15 & 1.436 & 0.0023 & 28.893 & 18.660 & 12.919   &  .16 & 1.588 & 0.0034 & 28.884 & 18.485 & 13.213 \\ 
.17 & 1.473 & 0.0031 & 28.855 & 18.426 & 13.202   &  .18 & 1.460 & 0.0026 & 28.868 & 18.471 & 13.570 \\ 
.19 & 1.512 & 0.0025 & 28.882 & 18.484 & 13.299   &  .20 & 1.464 & 0.0026 & 28.884 & 18.447 & 13.255 \\ 
.21 & 1.481 & 0.0030 & 28.871 & 18.491 & 13.184   &  .22 & 1.432 & 0.0023 & 28.888 & 18.511 & 13.073 \\ 
.23 & 1.470 & 0.0026 & 28.886 & 18.594 & 13.312   &  .24 & 1.471 & 0.0029 & 28.877 & 18.657 & 13.017 \\ 
.25 & 1.424 & 0.0034 & 28.868 & 18.609 & 13.019   &  .26 & 1.603 & 0.0025 & 28.885 & 18.595 & 12.996 \\ 
.27 & 1.611 & 0.0025 & 28.884 & 18.660 & 12.977   &  .28 & 1.483 & 0.0036 & 28.879 & 18.660 & 13.536 \\ 
.29 & 1.555 & 0.0027 & 28.879 & 18.580 & 13.348   &  .30 & 1.388 & 0.0026 & 28.874 & 18.758 & 13.037 \\ 
.31 & 1.410 & 0.0029 & 28.881 & 18.683 & 13.030   &  .32 & 1.313 & 0.0035 & 28.870 & 18.972 & 12.846 \\ 
\hline
\multicolumn{12}{l}{BLG501 ($V$-band)} \\ 
.01 & 1.482 & 0.0028 & 28.828 & 18.175 & 13.286   &  .02 & 1.406 & 0.0024 & 28.832 & 18.327 & 13.108 \\ 
.03 & 1.422 & 0.0027 & 28.856 & 18.343 & 13.515   &  .04 & 1.442 & 0.0025 & 28.866 & 18.406 & 13.186 \\ 
.05 & 1.516 & 0.0023 & 28.875 & 18.467 & 13.186   &  .06 & 1.491 & 0.0025 & 28.872 & 18.560 & 13.241 \\ 
.07 & 1.580 & 0.0035 & 28.876 & 18.620 & 13.363   &  .08 & 1.455 & 0.0030 & 28.839 & 17.965 & 13.481 \\ 
.09 & 1.419 & 0.0024 & 28.851 & 18.178 & 13.330   &  .10 & 1.410 & 0.0025 & 28.860 & 18.238 & 13.310 \\ 
.11 & 1.431 & 0.0024 & 28.869 & 18.358 & 13.068   &  .12 & 1.508 & 0.0032 & 28.870 & 18.492 & 13.383 \\ 
.13 & 1.438 & 0.0025 & 28.881 & 18.538 & 13.003   &  .14 & 1.493 & 0.0024 & 28.889 & 18.570 & 13.041 \\ 
.15 & 1.437 & 0.0025 & 28.888 & 18.698 & 12.919   &  .16 & 1.590 & 0.0040 & 28.880 & 18.476 & 13.213 \\ 
.17 & 1.460 & 0.0026 & 28.826 & 18.014 & 13.202   &  .18 & 1.460 & 0.0024 & 28.837 & 18.150 & 13.570 \\ 
.19 & 1.497 & 0.0024 & 28.853 & 18.244 & 13.299   &  .20 & 1.471 & 0.0025 & 28.867 & 18.275 & 13.255 \\ 
.21 & 1.490 & 0.0029 & 28.878 & 18.419 & 13.184   &  .22 & 1.451 & 0.0026 & 28.886 & 18.496 & 13.073 \\ 
.23 & 1.468 & 0.0028 & 28.885 & 18.585 & 13.312   &  .24 & 1.499 & 0.0027 & 28.899 & 18.618 & 13.017 \\ 
.25 & 1.431 & 0.0029 & 28.884 & 18.547 & 13.019   &  .26 & 1.538 & 0.0022 & 28.862 & 18.224 & 12.996 \\ 
.27 & 1.572 & 0.0016 & 28.870 & 18.415 & 12.977   &  .28 & 1.470 & 0.0026 & 28.872 & 18.531 & 13.536 \\ 
.29 & 1.540 & 0.0029 & 28.880 & 18.506 & 13.348   &  .30 & 1.410 & 0.0032 & 28.873 & 18.702 & 13.037 \\ 
.31 & 1.411 & 0.0027 & 28.885 & 18.598 & 13.030   &  .32 & 1.314 & 0.0032 & 28.885 & 18.892 & 12.846 \\ 
\hline
\multicolumn{12}{l}{BLG502 ($V$-band)} \\ 
.01 & 1.356 & 0.0041 & 28.791 & 18.362 & 13.286   &  .02 & 1.299 & 0.0038 & 28.797 & 18.549 & 13.108 \\ 
.03 & 1.311 & 0.0034 & 28.805 & 18.495 & 13.515   &  .04 & 1.307 & 0.0031 & 28.809 & 18.440 & 13.186 \\ 
.05 & 1.399 & 0.0032 & 28.818 & 18.480 & 13.186   &  .06 & 1.343 & 0.0027 & 28.810 & 18.511 & 13.241 \\ 
.07 & 1.395 & 0.0030 & 28.810 & 18.489 & 13.363   &  .08 & 1.288 & 0.0048 & 28.782 & 18.315 & 13.481 \\ 
.09 & 1.278 & 0.0038 & 28.790 & 18.537 & 13.330   &  .10 & 1.295 & 0.0035 & 28.800 & 18.588 & 13.310 \\ 
.11 & 1.326 & 0.0032 & 28.805 & 18.624 & 13.068   &  .12 & 1.419 & 0.0043 & 28.794 & 18.665 & 13.383 \\ 
.13 & 1.351 & 0.0037 & 28.808 & 18.600 & 13.003   &  .14 & 1.363 & 0.0023 & 28.813 & 18.507 & 13.041 \\ 
.15 & 1.297 & 0.0023 & 28.811 & 18.533 & 12.919   &  .16 & 1.387 & 0.0030 & 28.794 & 18.313 & 13.213 \\ 
.17 & 1.331 & 0.0044 & 28.773 & 18.416 & 13.202   &  .18 & 1.332 & 0.0035 & 28.794 & 18.484 & 13.570 \\ 
.19 & 1.399 & 0.0033 & 28.795 & 18.489 & 13.299   &  .20 & 1.360 & 0.0030 & 28.806 & 18.420 & 13.255 \\ 
.21 & 1.333 & 0.0034 & 28.798 & 18.465 & 13.184   &  .22 & 1.307 & 0.0025 & 28.803 & 18.484 & 13.073 \\ 
.23 & 1.314 & 0.0031 & 28.808 & 18.520 & 13.312   &  .24 & 1.351 & 0.0028 & 28.808 & 18.515 & 13.017 \\ 
.25 & 1.282 & 0.0027 & 28.783 & 18.393 & 13.019   &  .26 & 1.433 & 0.0032 & 28.799 & 18.471 & 12.996 \\ 
.27 & 1.463 & 0.0026 & 28.807 & 18.577 & 12.977   &  .28 & 1.343 & 0.0033 & 28.801 & 18.570 & 13.536 \\ 
.29 & 1.370 & 0.0039 & 28.798 & 18.499 & 13.348   &  .30 & 1.261 & 0.0026 & 28.789 & 18.632 & 13.037 \\ 
.31 & 1.273 & 0.0026 & 28.794 & 18.566 & 13.030   &  .32 & 1.206 & 0.0031 & 28.786 & 18.877 & 12.846 \\ 
\hline
\multicolumn{12}{l}{\ldots table abbreviated, see on-line material.} \\ 
\hline
\caption{%
The errorbar-correction coefficients for the $V$-band
data from the 1.3-m Warsaw Telescope made with the 
32-CCD-chip mosaic OGLE-IV camera.
The observing fields in the table are from the 
on-going microlensing survey toward
the Galactic bulge.
The full contents of this table is available 
on-line at \emph{ftp://ftp.astrouw.edu.pl/ogle/ogle4/errorbars/errcorr-OIV-BLG-V.dat}.
}
\end{longtable}
\end{footnotesize}


\begin{references}

\refitem{Alard, C., \& Lupton, R.~H.}{1998}{\apj}{503}{325}

\refitem{Bozza, V., Dominik, M., Rattenbury, N.~J., \etal}{2012}{\mnras}{424}{902}

\refitem{Dominik, M., Horne, K., Allan, A., \etal}{2008}{Astronomische Nachrichten}{329}{248}

\refitem{Gould, A., Yee, J., \& Carey, S.}{2015}{Spitzer Proposal}{12013}{}

\refitem{Mr{\'o}z, P., Udalski, A., Poleski, R., \etal}{2015a}{\apjs}{219}{26}

\refitem{Mr{\'o}z, P., Udalski, A., Poleski, R., \etal}{2015b}{\actaa}{65}{313}

\refitem{Pilecki, B., Graczyk, D., Pietrzy{\'n}ski, G., \etal}{2013}{\mnras}{436}{953}

\refitem{Pojma{\'n}ski, G.}{2002}{\actaa}{52}{397}

\refitem{Poleski, R., Udalski, A., Skowron, J., \etal}{2011}{\actaa}{61}{123}

\refitem{Skowron, J., Shin, I.-G., Udalski, A., \etal}{2015}{\apj}{804}{33}

\refitem{Skowron, J., Udalski, A., Poleski, R., \etal}{2016}{\apj}{820}{4}

\refitem{Smolec, R., \& Sniegowska, M.}{2016}{\mnras}{458}{3561}

\refitem{Soszy{\'n}ski, I., Udalski, A., Szyma{\'n}ski, M.~K., \etal}{2014}{\actaa}{64}{177}

\refitem{Sumi, T., Abe, F., Bond, I.~A., \etal}{2003}{\apj}{591}{204}


\refitem{Udalski, A., Szyma{\'n}ski, M., Ka{\l}u{\.z}ny, J., \etal}{1994}{\actaa}{44}{227}


\refitem{Udalski, A.}{2008}{\actaa}{58}{187}


\refitem{Udalski, A., Szyma{\'n}ski, M.~K., \& Szyma{\'n}ski, G.}{2015}{\actaa}{65}{1}

\refitem{Wo{\'z}niak, P.~R., Alard, C., Udalski, A., \etal}{2000}{\apj}{529}{88}

\refitem{Wo{\'z}niak, P.~R.}{2000}{\actaa}{50}{421}

\refitem{Wyrzykowski, {\L}., Koz{\l}owski, S., Skowron, J., \etal}{2009}{\mnras}{397}{1228}

\refitem{Wyrzykowski, L., Skowron, J., Koz{\l}owski, S., \etal}{2011}{\mnras}{416}{2949}

\refitem{Wyrzykowski, {\L}., Kostrzewa-Rutkowska, Z., Koz{\l}owski, S., \etal}{2014}{\actaa}{64}{197}

\end{references}
\end{document}